\documentclass{iaus}
\usepackage{graphicx}
\usepackage{natbib}
\def\hi{H{\sc i}}
\def\aap{A\&A}

\def\araa{Ann.\ Rev.\ A\&A}
\def\aj{AJ}
\def\apj{ApJ}
\def\apjl{ApJL}
\def\mnras{MNRAS}

\title[GHOSTS | Stellar halos of massive galaxies]{GHOSTS | Bulges, Halos, and the Resolved Stellar Outskirts of Massive Disk Galaxies}
\author[R.S.\ de Jong, D.J.\ Radburn-Smith \& J.N.\ Sick]{Roelof S.\ de Jong$^1$, David J.\ Radburn-Smith$^1$, and Jonathan N.\ Sick$^2$}
\affiliation{$^1$STScI, 3700 San Martin Dr., Baltimore, MD 21218, USA\\[\affilskip]
$^2$Rice University, Houston, TX 77005, USA}

\pubyear{2008}
\volume{254}  
\pagerange{1-6}
\jname{The Galaxy Disk in Cosmological Context}
\editors{J.~Andersen, J.~Bland-Hawthorn \& B.~Nordstr\"om, eds.}
\begin{document}

\maketitle

\begin{abstract}
  Our GHOSTS survey measures the stellar envelope properties of 14
  nearby disk galaxies by imaging their resolved stellar populations
  with HST/ACS\&WFPC2.  Most of the massive galaxies in the sample
  ($V_{\rm rot}$$>$200 km/s) have very extended stellar envelopes with
  S\'ersic law profiles or $\mu(r)$$\sim$$r^{-2.5}$ power law profiles
  in the outer regions. For these massive galaxies we can fit the
  central bulge light and the outer halo out to 30\,kpc with one and
  the same S\'ersic profile. The stellar surface density of these
  profiles correlate with Hubble type and bulge-to-disk ratio,
  suggesting that the central bulges and inner halos are created in
  the same process. Smaller galaxies ($V_{\rm rot}$$\sim$100 km/s)
  have much smaller stellar envelopes, but depending on geometry they
  could still be more luminous than expected from satellite remnants
  in hierarchical galaxy formation models. Alternatively, they could
  be created by disk heating through the bombardment of small dark
  matter sub-halos. The halos we fit are highly flattened, with
  minor-over-major axis ratios $c/a$\,$\simeq$\,0.4. The halos are
  somewhat more compact than hierarchical model predictions. The halos
  show small metallicity gradients out to 30\,kpc and the massive
  galaxies have typical [Fe/H]\,$\sim$\,$-0.8$. We find indications of
  halo substructure in many galaxies, but some halos seem remarkable
  smooth.

\end{abstract}

Hierarchical galaxy formation in a $\Lambda$CDM cosmology has become
the standard pa\-ra\-digm in recent years. However, our understanding of
the galaxy formation process is incomplete.
Which high redshift galaxy building blocks end up in what kind of
local galaxies? How much of the stellar content of the different
galaxy components (bulge, thin and thick disk, stellar halo) is
created in situ and how much is accreted? How does the current
accretion rate compare to $\Lambda$CDM predictions? To address these
questions we have begun the GHOSTS\footnote{GHOSTS: Galaxy Halos,
Outer disks, Substructure, Thick disks, and Star clusters} Survey,
using HST to perform stellar archaeology in the outskirts of 14 nearby
disk galaxies (8 of which are edge-on). For more details on the GHOSTS
survey see Radburn-Smith et al., in preparation.

We obtained HST/ACS observations in the F606W and F814W bands, with
typically 2--3 ACS pointings along the major and minor axes of each
galaxy.  Our observations reach 1.5 to 2 magnitudes below the
tip of the Red Giant Branch (RGB), allowing us to identify distinct
features in a Color-Magnitude Diagram (CMD) that relate to stellar
populations of very different ages \citep[for details
see][]{deJ07}. We can investigate the spatial distribution of stars in
each of these features to constrain formation histories of the
different galaxy components. Using this method we have already
constrained models of disk truncations in NGC\,4244 \citep{deJ07}.

\section{Bulge and halo surface density profiles}

We select RGB stars from our CMDs and use those to trace the stellar
surface density. RGB stars are ideal as they are abundant in our CMDs,
are indicative of old stellar populations (as expected to be found in
the outskirts of galaxies), and are representative of the underlying
stellar mass. To map the surface brightness profiles in the central
regions of the galaxies we use the integrated light from Spitzer/IRAC
4.5 micron observations. The Spitzer images provide near unobscured
light profiles, even for our edge-on galaxies. We scale the RGB
surface density star counts such that they match the IR luminosity
profiles in the overlapping region. In this way we derive equivalent
surface brightness profiles directly from the RGB star counts.

We find that all galaxies show extended components beyond the inner
exponential disk component. For the small and more face-on galaxies in
the sample the size and shape of the extended component is somewhat
unclear due to the uncertain contribution of contamination (unresolved
background galaxies, image defects, etc.) limiting the range where the
outer profile can be determined for these galaxies. Several of the
most massive galaxies have very extended envelopes with stellar
densities at 30 kpc that are 10--100 times higher than the
contamination background, with equivalent surface brightnesses of
about 29 $V$-mag arcsec$^{-2}$ at these radii.

\begin{figure}
\includegraphics[bb=83 207 500 550,clip=true,width=0.49\textwidth]{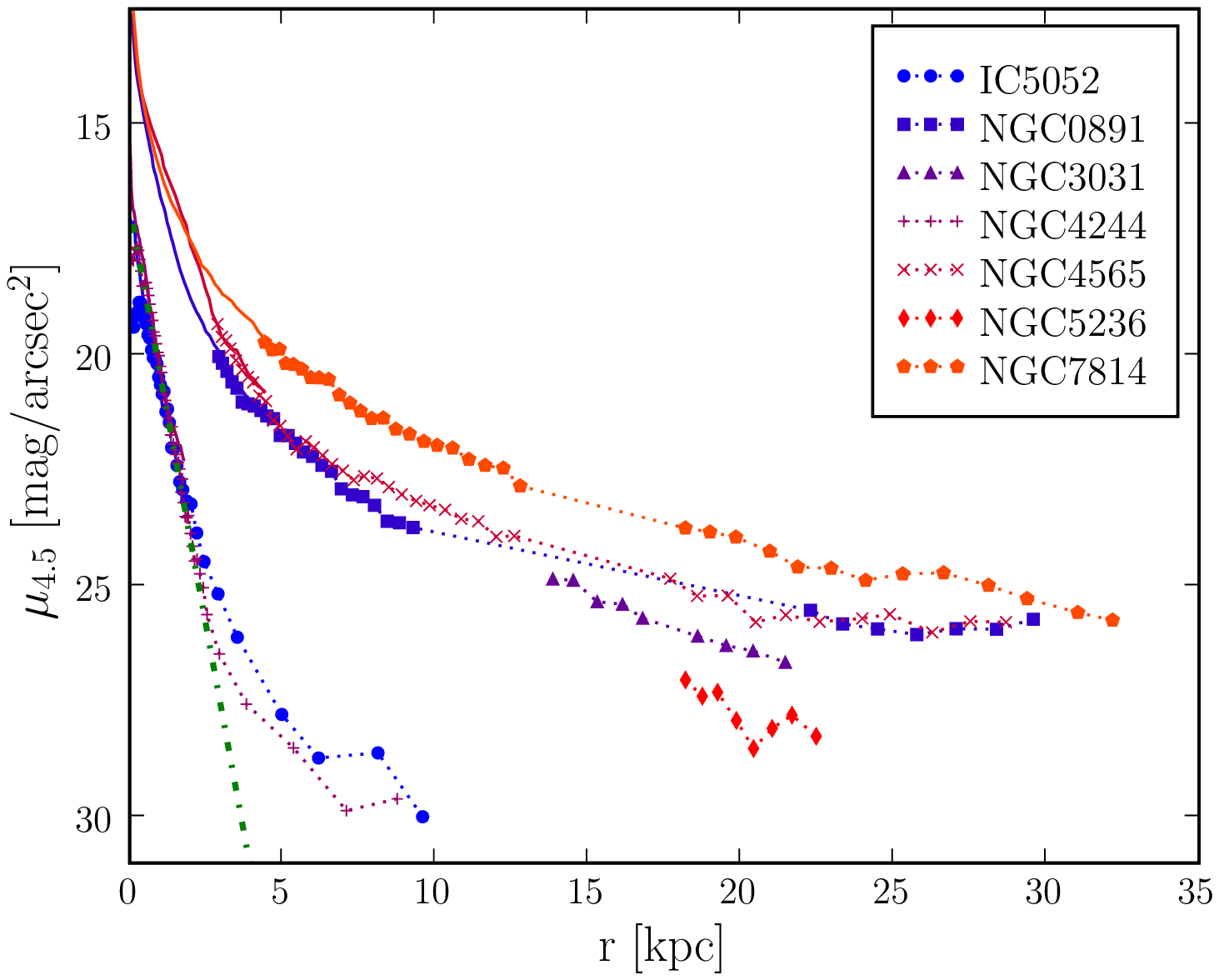}
\hfill
\includegraphics[bb=83 207 500 550,clip=true,width=0.49\textwidth]{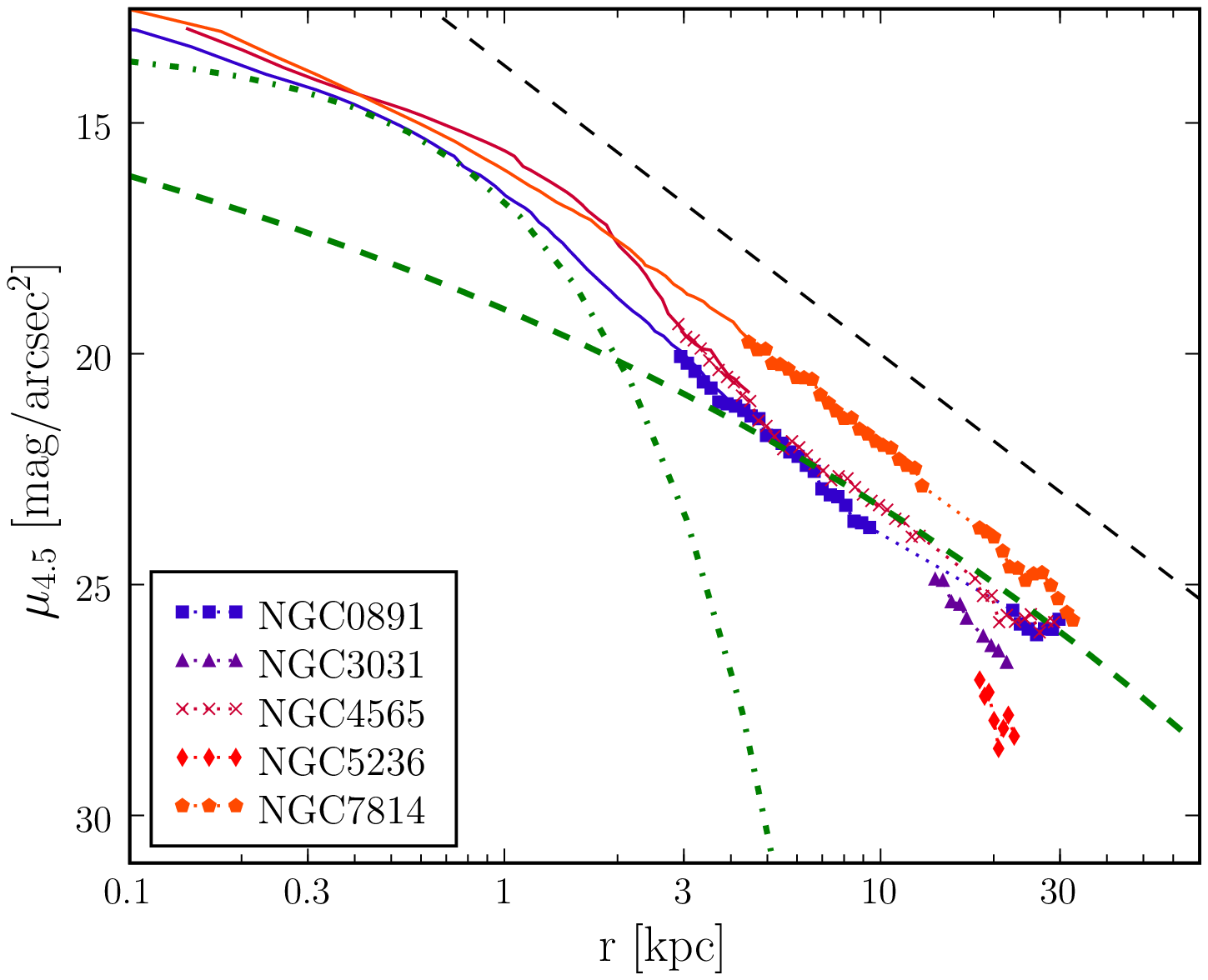}
\caption{Minor axis surface density profiles of GHOSTS galaxies. The
  thin solid lines indicate the profiles derived from Spitzer/IRAC 4.5
  micron images calibrated to Vega magnitudes (add about 3.5 mag to
  convert to Vega $V$-mag). The symbols connected with dotted lines
  represent RGB star count profiles, scaled to match the Spitzer
  data. To reduce confusion at small radii we only plot star counts
  beyond 12 kpc for the non-edge-on galaxies NGC\,3031/M81 and
  NGC\,5236/M83. On a linear radial scale (left diagram) exponential
  disks appear as straight lines, as indicated for IC5052 by the
  dot-dashed thick line. In the log-log plot on the right, where we
  have removed low mass galaxies for clarity, a straight line
  indicates a power law profile (e.g., thin dashed line =
  r$^{-2.5}$). Also shown are an exponential disk (for NGC\,0891,
  dot-dashed line) and a S\'ersic profile with the typical parameters
  for a flattened stellar halo as modeled by \citet{AbaNav06} (thick
  dashed line).  
\label{profs}
}
\end{figure}

In Fig.\,\ref{profs} we show minor-axis profiles of the edge-on
galaxies analyzed so far, along with outer profiles for the more
face-on galaxies NGC\,3031/M81 and NGC\,5236/M83%
. The exponential thin disks only dominate the inner $\sim$2--3 kpc
(5-10 scale heights), while the extended components are evident at
larger radii.  We find that all nine galaxies analyzed thus far show
components that are more extended than the exponential disks detected
at small radii.  Only for NGC\,5236/M83 is the evidence for an
extended component less certain. We measure a pure exponential disk
for this galaxy to at least 20 kpc (more than 10 disk-scale lengths),
but a field at 40\,kpc deviates strongly from the exponential
trend. However the stellar density of this outer field is close to the
sky background contamination. Furthermore, NGC\,5236/M83 has at least
two stellar streams at larger radii, indicating the presence of
significant substructure in the halo.

\subsection{The bulge-halo connection}
In this section we explore the connection between bulges and the
extended components. 

\begin{figure}[!tbh]
\includegraphics[height=0.24\textheight]{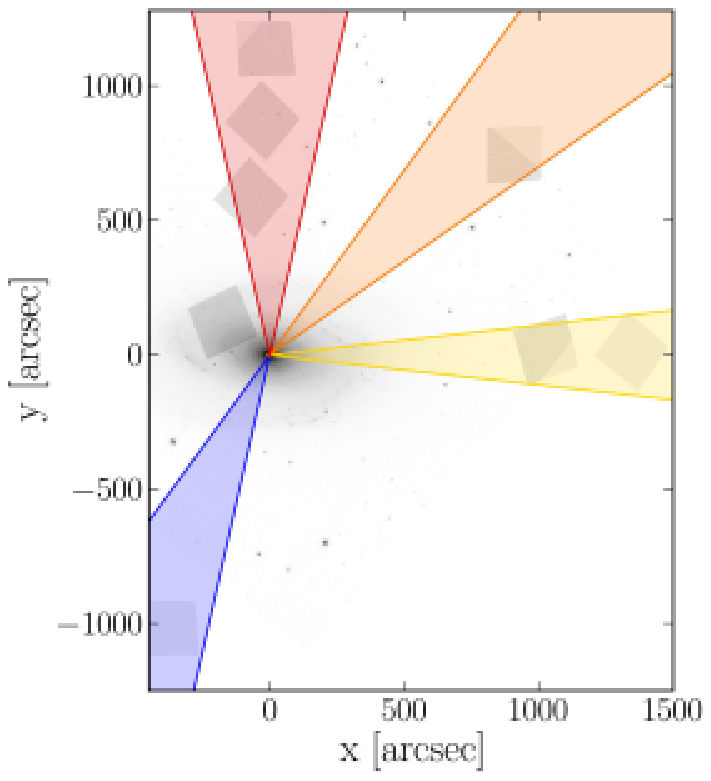}
\hfill
\includegraphics[width=0.50\textwidth]{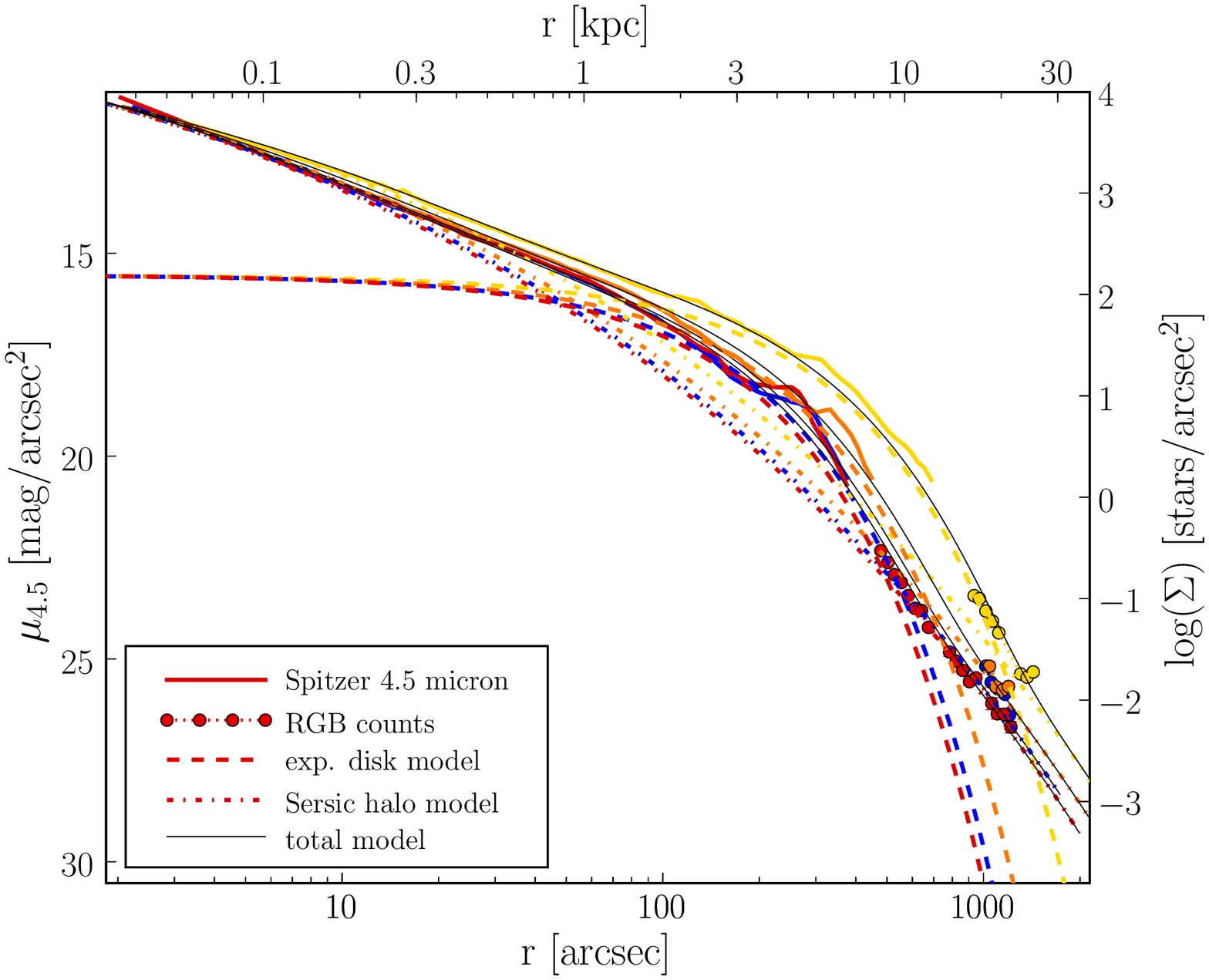}\\
\includegraphics[height=0.24\textheight]{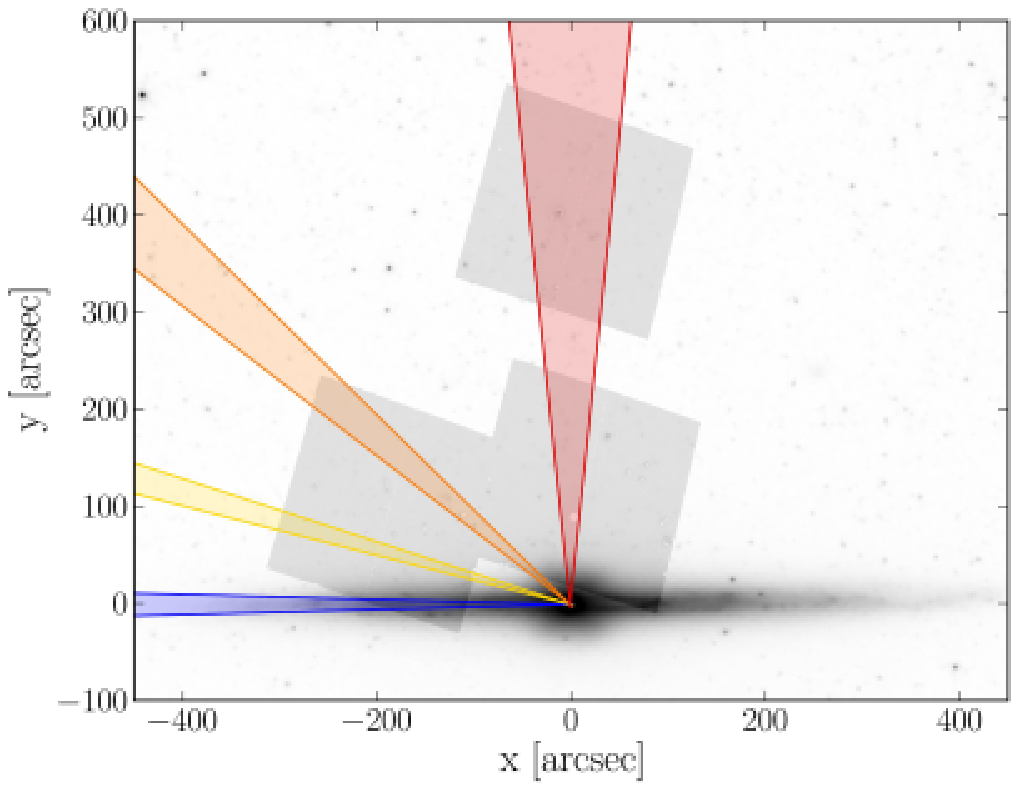}
\hfill
\includegraphics[width=0.50\textwidth]{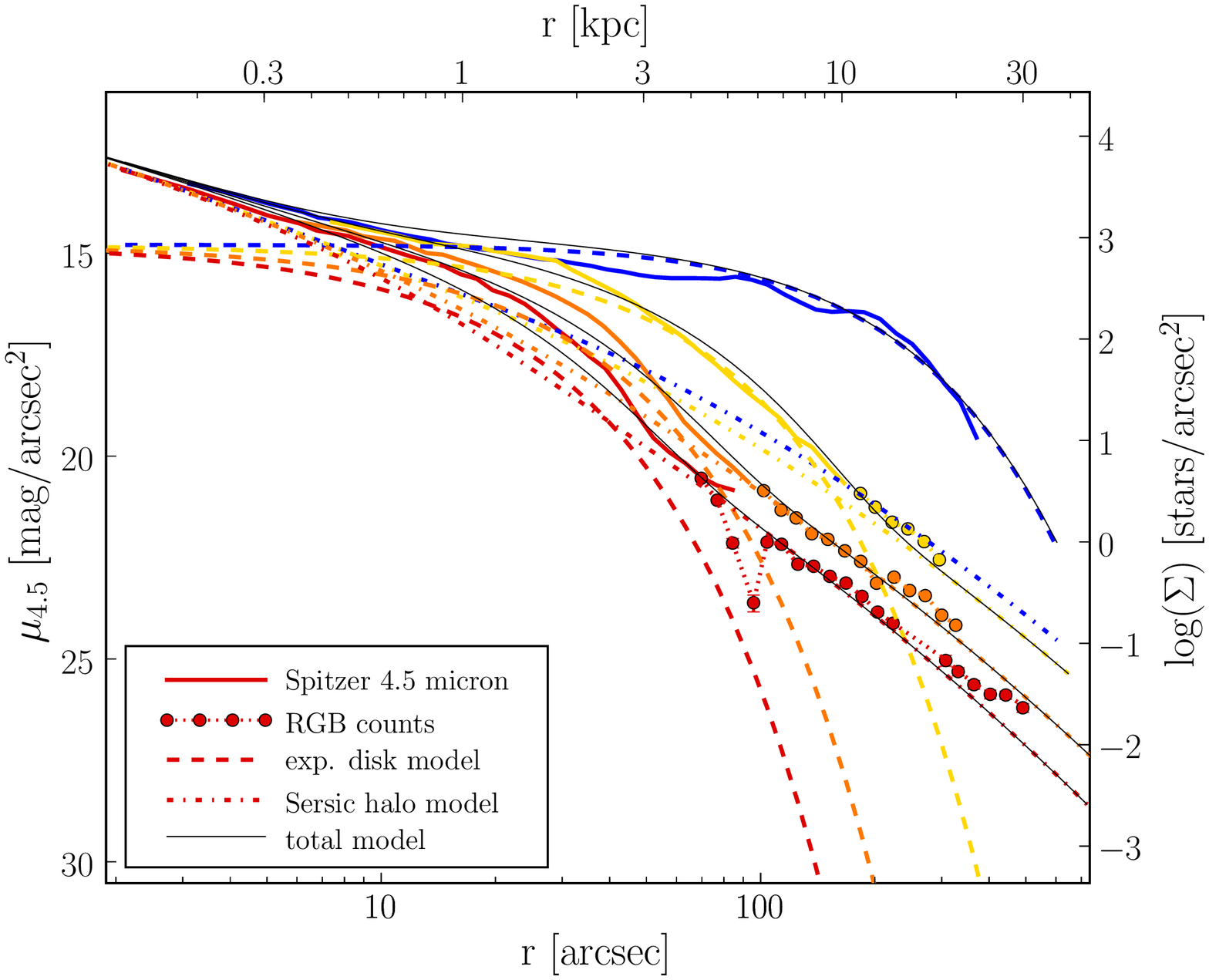}\\
\includegraphics[height=0.24\textheight]{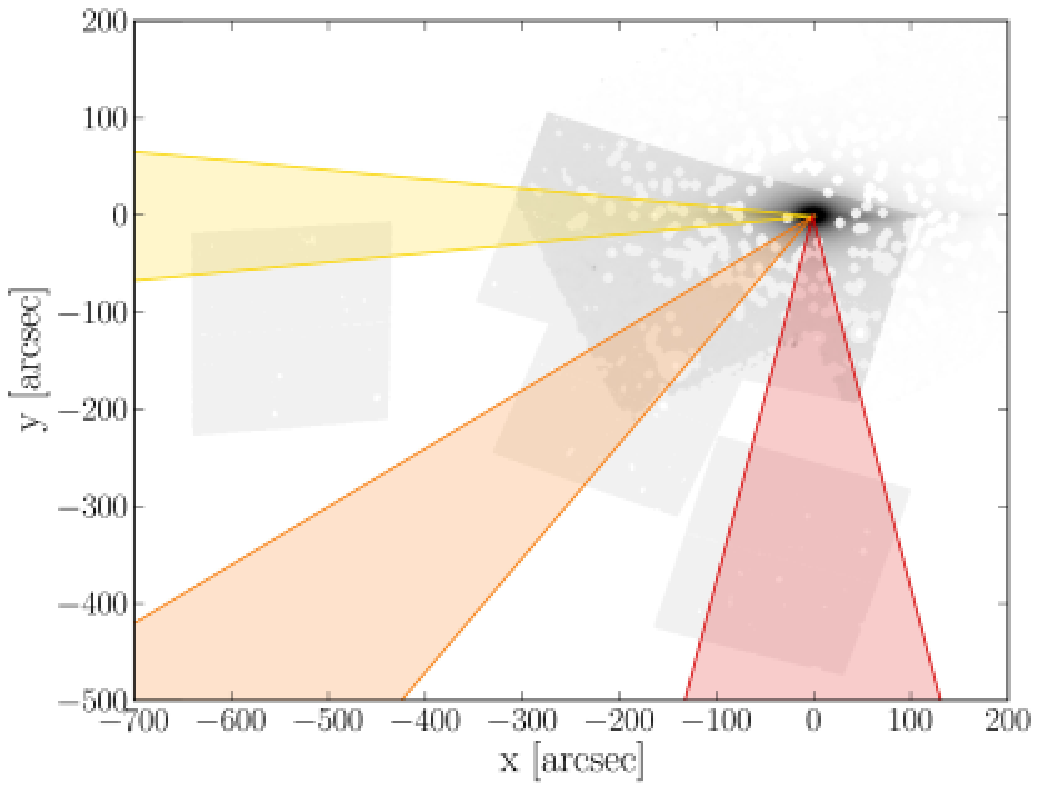}
\hfill
\includegraphics[width=0.50\textwidth]{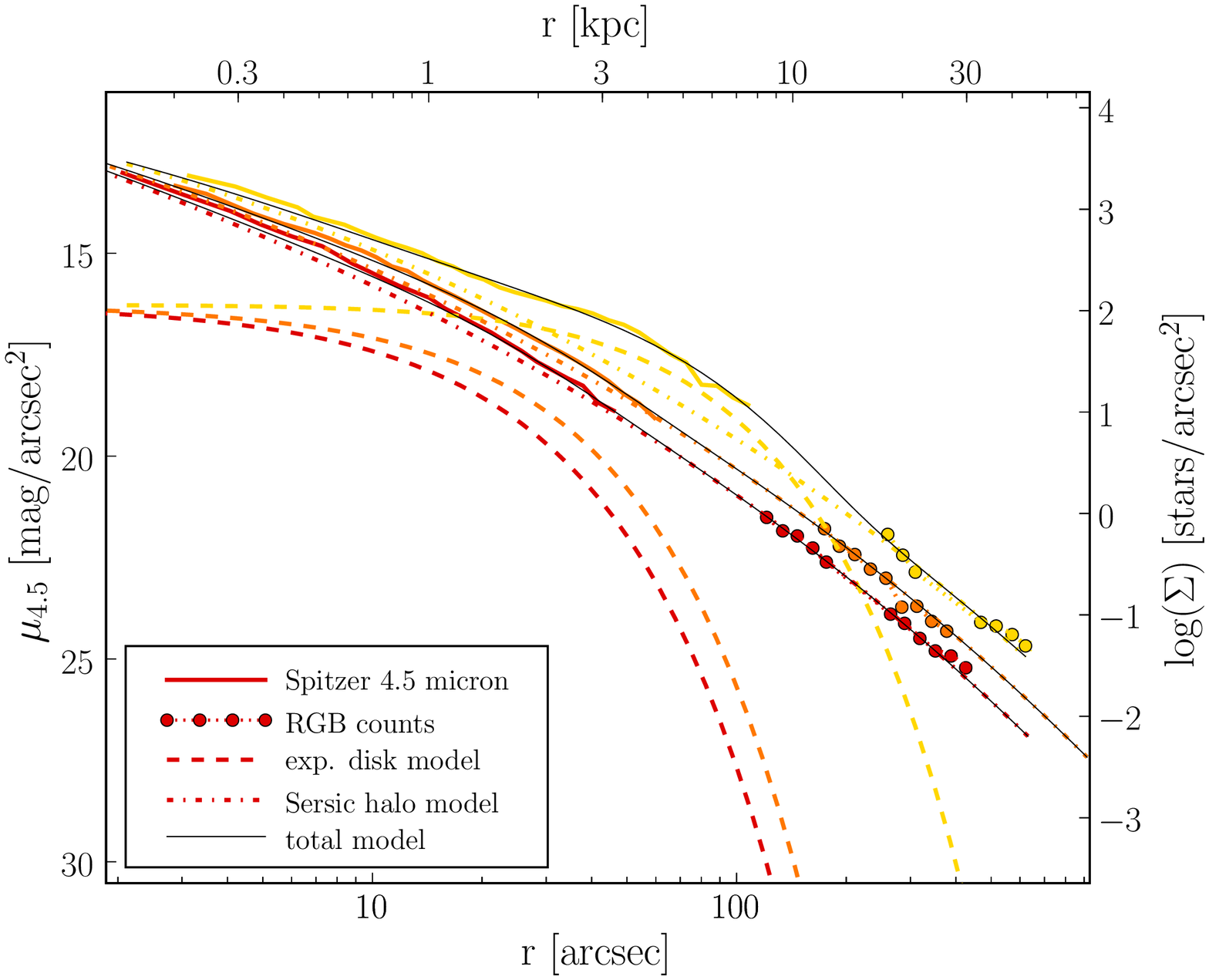}\\
\vspace*{-0mm}
\caption{
Surface density distribution and model fits of NGC\,3031, NGC\,4565,
and NGC\,7814 (top to bottom). \textit{Left:} Gray-scale images of the
{\em Spitzer} 4.5 micron images, with our HST/ACS footprints overlayed (shaded gray rectangular regions). The colored wedges indicate the regions used to
extract the surface brightness and stellar density profiles on the
right using the same color coding.  \textit{Right:} Radial {\em
  Spitzer} 4.5 micron surface brightness profiles (solid colored
lines, left axis) and matched RGB stellar surface density profiles
(points, right axis) color coded to the wedges in the diagrams to
the left. The best fitting galaxy component model is shown with the
exponential disk identified with long-dashed lines, the S\'ersic halo
with dot-dashed lines, and the combined model with thin solid lines,
again color coded for the different wedges.
\label{proffit}
}
\end{figure}

We fit two component 2D models to the inner 4.5 micron surface
brightness and outer star count data. The first component consists of
a projected disk with an exponential radial profile and a $sech$
vertical profile \citep{vdK88}. The second component is a flattened
halo with a S\'ersic radial density profile and a minor-over-major
axis ratio of $c/a$.  Merger models show that the hot components
resulting after a violent relaxation generally exhibit a S\'ersic
profile \citep[e.g.,][and reference therein]{BarHer92,AbaNav06}.  If
bulges and stellar envelopes are created by a collisionless merger
processes, we thus expect their light to follow a S\'ersic profile. We
show examples of the fits for a few galaxies in Fig.\,\ref{proffit}.

The profiles of the massive galaxies in our sample ($V_{\rm
  rot}$$>$200 km\,s$^{-1}$) can be fit entirely over a factor of 1000
in size ($\sim$10$^{4.5}$ in surface density) by an exponential disk
and a single S\'ersic distribution that represents both the inner
bulge-like region and the outer envelope. This simple model obviously
does not capture the fluctuations due to spiral arms, nor can it
describe the boxy (pseudo-)bulge in edge-on galaxies NGC\,891 and
NGC\,4565 (but does describe the central overdensity inside this boxy
pseudo-bulge). The fitted models are also unable to capture the halo
substructure seen as asymmetries and profile irregularities as
described in the next section. The fitted halos are highly flattened,
with typical $c/a$ values of 0.4 for the edge-on galaxies.

The smaller galaxies in the sample ($V_{\rm rot}$$\simeq$100--150
km\,s$^{-1}$) have small extended components, barely discernible above
the background contamination (see Fig.\,\ref{profs}). The shape of the
extended component is thus poorly constrained due to both the
uncertainty in the background and low number statistics. Furthermore,
the central bulge region provides no extra constraint as these
galaxies have no bulge, or at most contain a central star cluster. The
star counts can be fitted equally well by exponential, power law, and
S\'ersic law profiles. This extended feature could be the thick disk,
as observations along the NGC\,4244 major axis suggest the component
is very flattened. However, the disk as traced by RGB stars, whose
scale height is twice that of the main sequence population, has
already been identified as the thick disk \citep{Seth05II}, hence the
feature observed here is likely an additional component. These
additional components are most likely (depending on exact shape) more
luminous than predicted in the hierarchical models of
\citet{PurBul07}, but could have been created by the bombardment of
small dark matter sub-halos \citep{KazBul07}.

 
\subsection{Envelope properties and halo models}
Comparing the envelopes of the different galaxies we find that the two
small galaxies have much smaller extended components than the larger
galaxies, with surface densities that are lower relative to their
disks. 
The more massive galaxies in our sample are very similar in terms of
mass, luminosity, and scale size. Still, there is significant
variation in outer envelope properties. At 20 kpc NGC\,891, NGC\,4565,
and NGC\,7814 have power law profiles with a slope of about
-2.5. NGC\,3031/M81 has a steeper profile, while at 20 kpc NGC\,5236
is still dominated by the (face-on) disk. At first sight, the envelope
luminosity at 20 kpc seems correlated with Hubble type and
bulge-to-disk ratio, with the bulge dominated NGC\,7814 being the
brightest and the late-type spiral NGC\,5236 showing only an uncertain
hint of an envelope beyond 30\,kpc. Whilst M31 also fits this trend,
the Sab galaxy NGC\,3031/M81 does not, as a steeper and fainter
profile is evident at 20 kpc on one side of the galaxy. Then again,
the higher counts in another field at 20\,kpc, albeit on the other
side of M81, is consistent with the general trend. This observed
asymmetry may be due to the multiple interactions M81 is undergoing or
due to halo substructure.

In Fig.\,\ref{profs} we also show a typical profile from the
\citet{AbaNav06} model of accreted stars. Our profiles are somewhat
shallower and mostly fainter than these models between 10 and 30
kpc. While the surface density normalization may be somewhat uncertain
in the models, the shape is quite well constrained. It could be that
the true halos only dominate at even larger radii and the slope
becomes even shallower at larger radii. However, in hierarchical
galaxy formation the halo and ``classical'' bulge are formed by the
same merging process, so there is no reason to suspect a large
structural change between bulge and halo. The S\'ersic radii derived
for our combined envelope and bulge fits are typically eight times
smaller than those of \citet{AbaNav06}. However, a number of
simulation parameters affect the concentration of the accreted
halos. Suppresing star formation earlier, e.g. by pushing the epoch of
reionization to higher redshift, ensures that only the most massive
dark matter sub-halos contain stars, which yields steeper and fainter
envelopes \citep{BekChi05}. Alternatively, if the stars in the
accreted satellites sit deeper in the potential wells of their dark
matter sub-halos, they will then be tidally stripped closer to the
main galaxy, also resulting in more concentrated halos
\citep{BulJoh05}.

\section{Halo substructure}


We use Delaunay Tessellations to reconstruct the surface density
distributions in our fields (Sick \& de Jong, in prep.). Many galaxies
show indications of halo substructure, even with our sparse sampling
technique. In M83 we clearly detect a known stream that is dominated
by old stars and has a metallicity [Fe/H]$\sim$-0.6 (Seth \& de Jong,
in prep.). For both M81 and M83 we find significantly different
surface densities on opposite sides of the galaxies at the same
distance from the center. Many galaxies show deviations from smooth
radial profiles, most notably NGC\,891, where our outer minor axis
field is much higher than expected based on an extrapolation of the
inner halo profile. NGC\,4631 has an obvious over-density associated
with the neighboring galaxy NGC\,4627 just to the north, but there is
also a clear over-density to the northwest. This over-density, that is
seen in main sequence, AGB, and RGB stars, is potentially associated
with an \hi\ stream in this strongly interacting system.

In stark contrast to these clear signs of substructure, we find no
signs of any substructure surrounding the disk of NGC\,4565. When
comparing these density reconstructions with the substructure observed
around M31 on the same physical scale, the NGC\,4565 halo appears to
be much smoother. We are currently developing techniques to quantify
halo substructure.

\section{Envelope metallicities}

\begin{figure}
\centering
\includegraphics[width=0.8\textwidth]{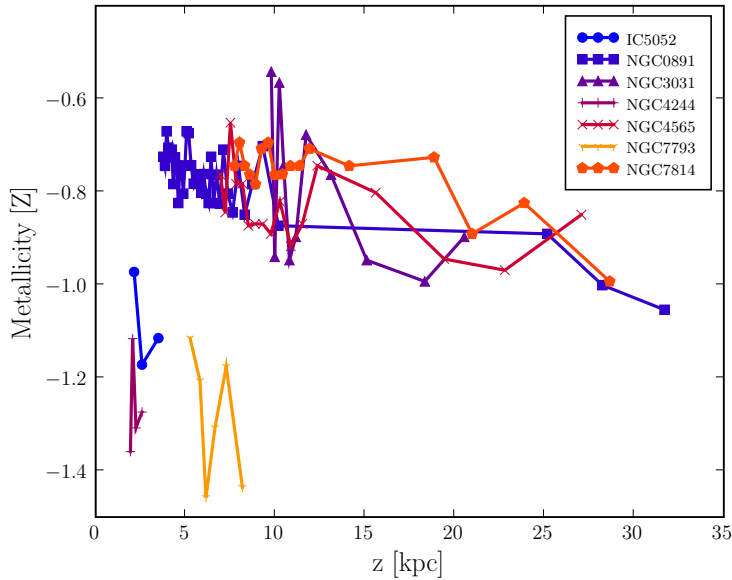}
\caption{
  Radial minor axis metallicity profiles derived from RGB star
  isochrone interpolation. We used isochrones of 10\,Gyr.
\label{radmet}
}
\end{figure}

The metallicities of the stellar envelopes can be derived from the
colors of the RGB stars, albeit with some degeneracy with population
age. We have used this technique to derive the radial metallicity
distribution for the GHOSTS galaxies assuming a typical population age
of 10\,Gyr (Fig.\,\ref{radmet}). Using younger isochrones would
increases the metallicity and correcting for the AGB contamination in
the RGB area will also increase the metallicity. The more massive
galaxies (NGC\,891, NGC\,3031, NGC\,4565, NGC\,7814) have
significantly higher metallicities than the lower mass galaxies
(IC5052, NGC4244, NGC7793) as has been seen before
\citep[e.g.,][]{Mou07}. Even though we detect halo metallicity
gradients for the massive galaxies, the metallicities at 30 kpc are
still much higher than found for the kinematically selected Milky Way
halo. We speculate that our flattened inner halos are the results of
more self-enriched, massive accreted satellites, while the hotter,
more metal poor halos may be detected at larger radii and are the
result of earlier, less enriched, smaller satellite accretion.

%

\acknowledgements 
Support for Proposal numbers 9765, 10523, and 10889 was provided by NASA
through a grant from the Space Telescope Science Institute, which is
operated by the Association of Universities for Research in Astronomy,
Incorporated, under NASA contract NAS5-26555.


\end{document}